\PassOptionsToPackage{usenames,dvipsnames}{xcolor}
\documentclass[sigconf]{acmart}
\AtBeginDocument{%
  \providecommand\BibTeX{{%
    \normalfont B\kern-0.5em{\scshape i\kern-0.25em b}\kern-0.8em\TeX}}}
\usepackage{dsfont}
\usepackage{times}
\usepackage{helvet}
\usepackage{courier}
\usepackage{dirtytalk}
\usepackage{graphicx} 
\usepackage{algorithm}
\usepackage{algorithmic}
\usepackage{graphicx}
\usepackage{subcaption}
\usepackage{mathtools}
\usepackage{amsthm,amsmath,amsfonts}
\usepackage{xspace}
\usepackage[margin=0.5cm]{subcaption}
\usepackage[capitalize,noabbrev]{cleveref}

\usepackage[textsize=tiny]{todonotes}

\usepackage{thmtools}

\DeclareUnicodeCharacter{0301}{\'{e}}

\newcommand{\cD}{\mathcal{D}}
\newcommand{\cE}{\mathcal{E}}

\newcommand{\E}[1]{\mathbb{E} \left[#1\right]}

\newcommand{\normw}[2]{\|#1\|_{#2}}

\newcommand{\set}[1]{\left\{#1\right\}}

\DeclareMathOperator*{\argmax}{arg\,max\,}

\mathchardef\mhyphen="2D

\copyrightyear{2022} 
\acmYear{2022} 
\setcopyright{rightsretained} 
\acmConference[WSDM '23]{The 16th ACM International Conference on Web Search and Data Mining}{February 27, March 3-23, 2023}{Singapore, Singapore}
\acmBooktitle{The 16th ACM International Conference on Web Search and Data Mining (WSDM '23), February 27, March 3, 2023, Singapore, Singapore}\acmDOI{xx.xxxx/xxxxxxxx.xxxxxxx}
\acmISBN{xxx-x-xxxx-xxxx-x/xx/xx}
\begin{document}

\title{From Ranked Lists to Carousels: A Carousel Click Model}

\hyphenation{re-com-men-der}

\sloppy

\author{Behnam Rahdari}
\affiliation{
  \institution{University of Pittsburgh}
}
\email{ber58@pitt.edu}

\author{Branislav Kveton}
\affiliation{
  \institution{Amazon}
}
\email{bkveton@amazon.com}

\author{Peter Brusilovsky}
\affiliation{
  \institution{University of Pittsburgh}
}
\email{peterb@pitt.edu}

\begin{abstract}
Carousel-based recommendation interfaces allow users to explore recommended items in a structured, efficient, and visually-appealing way. This made them a de-facto standard approach to recommending items to end users in many real-life recommenders. In this work, we try to explain the efficiency of carousel recommenders using a \emph{carousel click model}, a generative model of user interaction with carousel-based recommender interfaces. We study this model both analytically and empirically. Our analytical results show that the user can examine more items in the carousel click model than in a single ranked list, due to the structured way of browsing. These results are supported by a series of experiments, where we integrate the carousel click model with a recommender based on matrix factorization. We show that the combined recommender performs well on held-out test data, and leads to higher engagement with recommendations than a traditional single ranked list.
\end{abstract}


\maketitle

\section{Introduction}
\label{sec:introduction}

For many years, a ranked list of items was a de-facto standard for recommender systems to present recommendations and results to their users. Consequently, the majority of this research has been focused on developing and evaluating more and more powerful ranking algorithms. Following the example of information retrieval systems, recommender systems strive to accurately assess the relevance of candidate items and generate a ``perfect'' ranked list where the most relevant items are pushed to the top. Given this key goal, the power of a recommender system is evaluated by measuring its ability to estimate item relevance (i.e., predict item rating) and position it correctly in the ranked list. 
While relevance and ranking are still important issues in the field, an increasing volume of recommender systems research focuses on issues ``beyond ranking''. Most noticeable are several streams of research on ``collaborative'' recommendations where humans and AI work together to discover the most relevant items. It includes critiquing-based systems~\cite{chen12critiquingbased}, interactive recommender systems~\cite{chen2016interactive}, and user-controlled recommenders~\cite{jannach2017user}. Several types of \say{interactive} recommender interfaces where human and AI-based recommender system can collaborate in guiding users to the right items have been explored and their effectiveness have been convincingly demonstrated~\cite{smyth2004compound,bostandjiev2012tasteweights,vig2012tag,parra2015user}. Many projects also explore opportunities to present results of recommendation not as a one-dimensional ranked list, but as a two-dimensional grid~\cite{liang2021toward} or more complex visualization~\cite{verbert2016agents,kunkel20173D}

An interesting example of a recommender interface that combines the aspects of interactivity and 2D presentation of results is a carousel-based interface that is rapidly replacing the ranked list as a de-facto standard to present recommendations to end-users in e-commerce systems.
While the interface with multiple carousels (sometimes referred to as a \emph{multilist}) looks relatively complex -- it presents several ranked lists, each marked with a category, in place of a single ranked list -- it was embraced by the end-users, and industry. From the prospect of recommender systems, the carousel-based interface provides an excellent example of human-AI collaboration in the recommendation context. While a single ranked list attempts to be ``perfect'', in reality, the intent of the user is frequently uncertain. Most importantly, in many real-life applications users might have multiple interests and recommender systems rarely know which specific interest (for example, a movie genre) the users want to pursue at the given moment. A carousel-based interface leaves the task of choosing the most timely topic of interest (i.e., British documentaries) to the users. As a result, a user could quickly locate a ranked sub-list of the most relevant items, while also indirectly informing the recommender system about the kind of items they prefer right now. 

The popularity of carousel-based interfaces has not been ignored by the researchers on recommender systems. A growing number of papers focused on carousel-based interfaces have been published over the last few years~\cite{carousel1,gruson2019offline,felicioni2021methodology,jannach2021exploring,starke2021serving,liang2021toward}. Surprisingly, there are almost no attempts to evaluate carousel-based recommendations more formally and compare the value of carousel interfaces with traditional ranked lists. 

One of the issues that make it hard to compare carousel-based and ranked-list interfaces is the challenge to place them on equal ground for a fair evaluation. Traditional metrics used to compare ranking approaches such as Normalized Discounted Cumulative Gain (NDCG)~\cite{jarvelin2002cumulated} or Mean Reciprocal Rank (MRR)~\cite{mrr_Craswell2009} cannot be used directly for evaluating carousel-based interfaces. While some attempts to create metrics for evaluating multi-list presentations have been made~\cite{felicioni2021methodology}, a presentation-focused evaluation ignores user interaction with carousels, which is one of the keys to their power. In this situation, a fair opportunity to compare carousels with ranked lists could be offered by a simulation-based evaluation approach. The idea of this approach is a continuous simulation of user interactions with the systems being compared while computing various performance metrics ``on the go''. It could be applied to relatively complex interaction scenarios as long as user behavior in these scenarios could be modeled sufficiently well. Extensive studies of user interactions with a ranked list~\cite{granka2004eye,guan2007eye,keane2008are} helped to create a range of \emph{click models} that could be used for a reliable simulation of user work with ranked lists. However, no comparable click models of user interaction with carousel-based interfaces have been developed so far. 

Our work attempts to bridge this gap. The focus of this paper is a novel carousel click model that can simulate how users interact with topic-labeled carousel interfaces. In the following sections, we explain this model and present the results of several studies, which justify the model and demonstrate its use for assessing the value of a specific carousel interface against a ranked list alternative.

\section{Related Work}
\label{sec:background}

This section reviews related work in the areas of click models and simulation-based studies.

\subsection{Click Models in Recommender Systems}
The goal of \emph{click models} is to model and explain user interaction with a ranked list of results. This research stream started in the field of information retrieval and was originally motivated by the need to improve Web search engine performance by applying user \emph{click-through data} accumulated by search engine logs~\citep{zhu2001pagerate,sullivan2001how}. While ``old school'' information retrieval considered item relevance as the only factor determining user decision to click on a specific result, it became evident that the position of items in a ranked list has to be considered as well~\citep{joachims2002optimizing}. Moreover, creative experiments demonstrated that a high item position in a ranked list could provide a larger impact on click probability than true item relevance~\citep{keane2008are}. A sequence of eye-tracking studies with users of search engines~\citep{granka2004eye,pan2004determinants,guan2007eye,guan2007what} helped understand how users process a ranked list of results and measure the impact of item position in the list on the click probability. 

This research provided a solid ground for developing click models for user
interaction with ranked lists~\citep{chuklin15click}, which is now actively used in both information retrieval and recommender systems research. Many click models exist \citep{agichtein06learning,richardson07predicting,craswell08experimental,guo09click,guo09efficient,chapelle09dynamic}. Essentially all of them try to explain the user behavior by a generative model, which can be learned from data. As an example, the cascade model \citep{richardson07predicting,craswell08experimental} assumes that the user examines the list of recommended items from top to bottom until they find an attractive item. After that, they click on that item and leave satisfied. This seemingly simple model explains the observed position bias in recommender systems, that lower-ranked items are less likely to be clicked than higher-ranked items. This information can be used to debias click-through data \citep{li18offline}, or to learn better ranking policies either offline \citep{chuklin15click} or online \citep{kveton15cascading,combes15learning}. In this work, we study a generative model of user behavior that explains why interactions with carousels can be more efficient than with a single ranked list.

\subsection{Simulation-Based Studies of Recommender Systems}
\label{sec:models}
A simulation-based approach has been used for exploration and evaluation in a number of fields where sufficiently detailed models of user behavior could be built. In particular, a simulation-based study is a recognized approach for evaluating various kinds of personalized interactive systems from adaptive learning systems~\cite{champaign2013ecological,Drachsler2008} to personalized information access systems~\cite{snif-act2003,Mostafa2003}. The goals of simulation-based evaluation differ between application areas and frequently depend on the reliability of behavior models that support the simulation. On one end of the spectrum are cognitively grounded behavior models that are supported by studies of human cognition and confirmed by empirical studies. A well-known example is SNIF-ACT model~\cite{snif-act2003} that simulates user behavior in hypertext navigation. This model is based on Information Scent theory~\cite{card2001information} and was used to assess the quality and navigability of Web sites without real users. Popular ``simulated student'' models~\cite{matsuda2007predicting,champaign2013ecological} used for evaluation of adaptive educational systems also belong to this group. On the other end, there are a range of simple behavior models~\cite{yao2020fates} that might not be able to reliably predict the details of user behavior but could be useful to explore a range of ``what if'' scenarios in assessing the impact of various interface augmentations.   
 
Early attempts to use simulations for exploring information filtering and recommender systems were made in the first decade of 2000~\cite{Mostafa2003,Drachsler2008}. However, it took another 10 years for this approach to become truly noticed and used in this field~\cite{Dzyabura_2013,Hazrati_2022}. While the role of simulation-based research in the recommender system context is currently recognized, simulations are most frequently used for exploring the impact of recommender systems on various aspects of user behavior rather than assessing their performance and effectiveness in a comparative way. The most popular research direction enabled by simulation is examining the impact of a recommender system on various aspects of user behavior as a whole \cite{Drachsler2008,Bountouridis_2019,Hazrati_2020,yao2020fates}. This work is typically enabled by the user choice models~\cite{Hazrati_2022}. While research on click models reviewed in \cref{sec:models} offer a solid ground for simulation-based studies, there are very few cases where models of user click behavior were used for comparative off-line evaluation of recommender system design options.  A notable exception is the work of Dzyabura and Tuzhilin~\cite{Dzyabura_2013} who used simulation to compare an interface based on a combination of search and recommendation to interfaces based on search or recommendation alone. However, this work used a relatively simple behavior model that was not based on empirical observations or theory. In our work, we perform simulation-based studies using more complex and empirically grounded models, which increase our chances to obtain useful and reliable results. 

\section{Beyond Cascade Model}
\label{sec:beyond cascade model}

This section introduces the cascade click model and generalizes it to users that leave unsatisfied after examining unattractive items. The latter leads to our carousel click model in \cref{sec:carousel click model}. Our click models are defined over a ground set of items $E$ that could be recommended. The list of $K$ recommended items is a vector $A = (A_k)_{k = 1}^K$, where $A_k \in E $ is the item at position $k$. We start with the cascade model.

\subsection{Cascade Model}
\label{sec:cascade model}

The \emph{cascade model (CM)} \citep{richardson07predicting,craswell08experimental} is a popular model of user behavior in a ranked list. In this model, the user is recommended a list of $K$ items. The user examines the list from the first item to the last, and clicks on the first attractive item in the list. The items before the first attractive item are not attractive, because the user examines them but does not click on them. The items after the first attractive item are unobserved because the user never examines them.

The preference of the user for item $a \in E$ in the cascade model is represented by its \emph{attraction probability} $p_a \in [0, 1]$. The attraction of item $a$ is a random variable defined as $Y_a \sim \mathrm{Ber}(p_a)$. Fix list $A$. The click on position $k$ is denoted by $C_k$ and defined as $C_k = E_k Y_{A_k}$, where $E_k$ is an indicator that position $k$ is examined. By definition, the position is examined only if none of the higher-ranked items is attractive. Thus $E_k = \prod_{\ell = 1}^{k - 1} (1 - Y_{A_\ell})$ and the probability of a click on position $k$ is
\begin{align*}
  P_\textsc{cm}(A, k)
  = \E{E_k Y_{A_k}}
  = \left[\prod_{\ell = 1}^{k - 1} (1 - p_{A_\ell})\right] p_{A_k}\,.
\end{align*}
In turn, the probability of a click on list $A$ is
\begin{align}
  P_\textsc{cm}(A)
  = \E{\sum_{k = 1}^K E_k Y_{A_k}}
  = \sum_{k = 1}^K P_\textsc{cm}(A, k)\,.
  \label{eq:cm click probability}
\end{align}

This model has two notable properties. First, since \eqref{eq:cm click probability} increases whenever a less attractive item in $A$ is replaced with a more attractive item from $E$, the optimal list in the CM,
\begin{align*}
  \textstyle
  A_*
  = \argmax_A P_\textsc{cm}(A)\,,
\end{align*}
contains $K$ most attractive items. Therefore, it can be easily computed. Second, the click probability $P_\textsc{cm}(A, k)$ can be used to assess if a model reflects the ground truth. In particular, let $\hat{\mu} \in [0, 1]^K$ be the frequency of observed clicks on all positions in list $A$. Then the click model is a good model of reality if $\hat{\mu}$ resembles the output of the model. This similarity can be measured in many ways and we adopt the \emph{total variation distance} of click probabilities, $\frac{1}{2} \normw{P_\textsc{cm}(A, \cdot) - \hat{\mu}}{1}$, in this work.

A click model is a mapping from a ranked list to its relevance under a model of user behavior. For a click model $M$ and list $A$, let $P_M(A)$ be the relevance of list $A$ under model $M$, such as $P_\textsc{cm}(A)$ in \eqref{eq:cm click probability}. Click models can be used to answer several types of queries. The computation of $P_M(A)$ is \emph{evaluation} of the relevance of list $A$ under model $M$. A natural extension is a \emph{comparison} of two lists under a fixed model. Specifically, if $P_M(A) > P_M(A')$ for lists $A$ and $A'$, we may conclude that list $A$ is more relevant than list $A'$ under model $M$. Finally, we can also compare the same list under two different models. In particular, if $P_M(A) > P_{M'}(A)$ for models $M$ and $M'$, we may conclude that the user is more engaged with list $A$ under model $M$ than $M'$.

\subsection{Terminating Cascade Model}
\label{sec:terminating cascade model}

One shortcoming of the cascade model is that the order of items in the optimal list does not affect the click probability. This is why extending this model to structured problems is difficult, because the position of the item does not matter. To introduce dependence on the order of items, we modify the CM as follows. When the examined item is not attractive, the user leaves unsatisfied with \emph{termination probability} $p_q \in [0, 1]$. This models a situation where the user gets tired after examining unsatisfactory items. We call this model a \emph{terminating cascade model (TCM)}.

\begin{figure}[h!]
\centering
            \includegraphics[width=.90\textwidth/2]{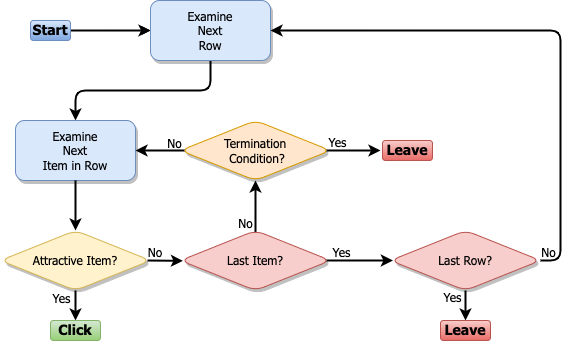}
            \caption[]%
            {{\small Schematic view of TCM.}}    
            \label{fig:tcm_schema}
 \end{figure}

Fix list $A$. Let $Q_k$ be an indicator that the user leaves unsatisfied at position $k$, which is defined as $Q_k \sim \mathrm{Ber}(p_q)$. Then the click on position $k$ is defined as $C_k = E_k Y_{A_k}$, where $E_k$ is an indicator that position $k$ is examined. Since the position is examined only if none of the higher-ranked items is attractive, and the user does not leave unsatisfied upon examining these items, we have
\begin{align*}
  E_k
  = \prod_{\ell = 1}^{k - 1} (1 - Q_\ell) (1 - Y_{A_\ell})\,.
\end{align*}
Thus the probability of a click on position $k$ is
\begin{align}
  P_\textsc{tcm}(A, k)
  = \E{E_k Y_{A_k}}
  = (1 - p_q)^{k - 1} \left[\prod_{\ell = 1}^{k - 1} (1 - p_{A_\ell})\right] p_{A_k}
  \label{eq:tcm position click probability}
\end{align}
and that on the list $A$ is
\begin{align}
  P_\textsc{tcm}(A)
  = \E{\sum_{k = 1}^K E_k Y_{A_k}}
  = \sum_{k = 1}^K P_\textsc{tcm}(A, k)\,.
  \label{eq:tcm click probability}
\end{align}
This model behaves similarly to the CM and has all of its desired properties. First, the optimal list in the TCM,
\begin{align*}
  \textstyle
  A_*
  = \argmax_A P_\textsc{tcm}(A)\,,
\end{align*}
contains $K$ most attractive items in descending order of their attraction probabilities. The order matters because the position $k$ in \eqref{eq:tcm click probability} is discounted by $(1 - p_q)^{k - 1}$. Interestingly, this list is invariant to the value of the termination probability, as long as $p_q \in (0, 1)$. Second, since $P_\textsc{tcm}(A, k)$ can be easily computed for any list $A$ and position $k$, the fit of the TCM to empirical click probabilities can be evaluated as in \cref{sec:cascade model}, using the total variation distance.

\section{Carousel Click Model}
\label{sec:carousel click model}

A natural way of extending the cascade model to carousels is to think of the recommended items as a matrix $A = (A_{i, j})_{i \in [m], j \in [K]}$, where $m$ is the number of carousels (rows), $K$ is the number of items per carousel (columns), and $A_{i, j}$ is the recommended item at position $(i, j)$. To simplify notation, we denote carousel $i$ in matrix $A$ by $A_{i, :} = (A_{i, j})_{j = 1}^K$. We assume that no item is in more than one carousel, that is $A_{i, j} \neq A_{i', j'}$ for any $(i, j) \neq (i', j')$.

The user examines the recommended matrix $A$ from the first carousel until the last. When \emph{carousel $i$ is attractive}, at least one item in $A_{i, :}$ is attractive, the user starts examining it and clicks on the first attractive item in it. When \emph{carousel $i$ is unattractive}, no item in $A_{i, :}$ is attractive, the user proceeds to the next carousel $i + 1$. When the user examines an unattractive carousel or item, they leave unsatisfied with probability $p_q \in [0, 1]$, similarly to the terminating cascade model in \cref{sec:terminating cascade model}. We call this model a \emph{carousel click model (CCM)}. 

\begin{figure}[h!]
\centering
            \includegraphics[width=.90\textwidth/2]{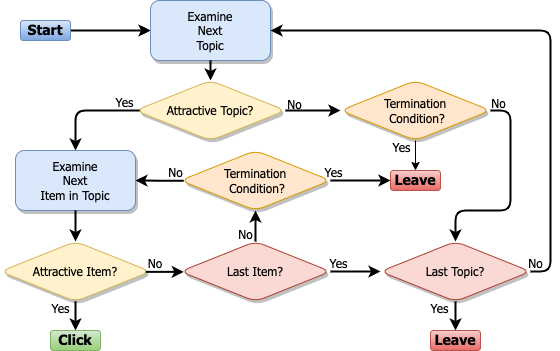}
            \caption[]%
            {{\small Schematic view of CCM.}}    
            \label{fig:ccm_schema}
 \end{figure}

The limiting factor of carousels is that the items in each carousel need to be semantically related. This amounts to a constraint on which items can be in $A_{i, :}$. The semantics also needs to be surfaced to the user, for instance by labeling the carousels with topics that the user can recognize. Indeed, only then the user could examine the carousel assuming that some items are attractive, as this would be suggested by an attractive topic associated with the carousel. This is the key assumption in our examination model.

\subsection{Click Probability}
\label{sec:click probability}

Fix matrix $A$. Let $Q_i \sim \mathrm{Ber}(p_q)$ be an indicator that the user leaves unsatisfied after examining carousel $i$. Let $Q_{i, j} \sim \mathrm{Ber}(p_q)$ be an indicator that the user leaves unsatisfied after examining item at position $(i, j)$. Then the indicator of a click on matrix $A$ can be written as
\begin{align*}
  \sum_{i = 1}^m E_i \sum_{j = 1}^K
  \left[\prod_{\ell = 1}^{j - 1} (1 - Q_{i, \ell}) (1 - Y_{A_{i, \ell}})\right] Y_{A_{i, j}}\,,
\end{align*}
where
\begin{align*}
  E_i
  = \prod_{\ell = 1}^{i - 1} (1 - Q_\ell) \prod_{j = 1}^K (1 - Y_{A_{\ell, j}})
\end{align*}
is an indicator that carousel $i$ is examined. The algebraic form of $E_i$ follows from the fact that even $E_i$ can occur only if all higher-ranked carousels are unattractive and the user does not leave unsatisfied after examining them. Thus the probability of a click on matrix $A$ is
\begin{align}
  P_\textsc{ccm}(A)
  = \sum_{i = 1}^m \cE_i P_\textsc{tcm}(A_{i, :})\,,
  \label{eq:ccm click probability}
\end{align}
where
\begin{align}
  \cE_i
  = (1 - p_q)^{i - 1} \prod_{\ell = 1}^{i - 1} \prod_{j = 1}^K (1 - p_{A_{\ell, j}})
  \label{eq:carousel examination probability}
\end{align}
is the probability that carousel $i$ is examined.

\subsection{Empirical Fit}

Similarly to the CM and TCM, we also have a closed form for the click probability on position $(i, j)$,
\begin{align}
  P_\textsc{ccm}(A, (i, j))
  = (1 - p_q)^{i + j - 2} \left[\sum_{\ell, s: (\ell < i) \vee (s < j)} \hspace{-0.15in}
  (1 - p_{A_{\ell, s}})\right] p_{A_{i, j}}\,.
  \label{eq:ccm position click probability}
\end{align}
This can be used to assess if a model reflects the ground truth. In particular, let $\hat{\mu} \in [0, 1]^{m \times K}$ be the frequency of observed clicks on all entries of matrix $A$. Then, if we treat $P_\textsc{ccm}(A, \cdot)$ and $\hat{\mu}$ as vectors, the total variation distance of click probabilities $\frac{1}{2} \normw{P_\textsc{ccm}(A, \cdot) - \hat{\mu}}{1}$ measures if the CCM is a good model of reality.

\subsection{Optimal Solution}
\label{sec:optimal solution}

The optimal solution in the CCM does not have a closed form anymore. Nevertheless, we can still characterize some of its properties. Specifically, in the optimal matrix $A_* = \argmax_A P_\textsc{ccm}(A)$, the items in each carousel must be ordered from the highest attraction probability to the lowest. This can be seen as follows. For any matrix $A$ and carousel $i$ in it, $P_\textsc{tcm}(A_{i, :})$ in \eqref{eq:ccm click probability} has the highest value when the attraction probabilities in $A_{i, :}$ are in a descending order. This argument is analogous to that in the TCM (\cref{sec:terminating cascade model}). This change has no impact on $\cE_i$ in \eqref{eq:carousel examination probability}. Regarding the order of carousels, we approximate $A_*$ by presenting the carousels in the descending order of their total attraction probabilities, $\sum_{j = 1}^K p_{A_{i, j}}$. This guarantees that carousels with more attractive items are presented first, which minimizes the probability of users leaving unsatisfied.

\subsection{Comparison to the TCM}
\label{sec:comparison tcm}

This section shows that the carousel click model can lead to higher click probabilities than the TCM, under the assumption that the attraction and termination probabilities in both models are comparable. Because the parameters of the models are comparable, this shows that the structure can be beneficial in recommendations.

We compare the TCM and CCM under the assumption that all attraction probabilities are identical and small. Specifically, let $p_a = p$ for all items $a$ and $p = O(1 / K m)$. Then $(1 - p)^k = O(1)$ for any $k \in [K m]$. In the TCM, we view $A$ as a single ranked list of $K m$ items. Then
\begin{align*}
  P_\textsc{tcm}(A)
  & \approx p \sum_{k = 1}^{K m} (1 - p_q)^{k - 1}\,, \\
  P_\textsc{ccm}(A)
  & \approx p \sum_{i = 1}^m \sum_{j = 1}^K (1 - p_q)^{i + j - 2}\,. 
\end{align*}
Now when we bound $j - 1$ in $P_\textsc{ccm}(A)$ as $j - 1 \leq K (j - 1)$, the two objectives become equal. Since $1 - p_q \leq 1$ and $j - 1 \geq 0$, we get $P_\textsc{tcm}(A) \leq P_\textsc{ccm}(A)$. The improvement is due to the fact that the user is much less likely to leave unsatisfied with the CCM.

\section{Experiments}
\label{sec:experiments}

To validate our model, we conduct four experiments. First, we evaluate \textsc{ccm} on a real-world dataset of carousel usage. Then, we validate the generalizability of our model. Next, we highlight the effect of personalization by comparing \textsc{ccm} to two similar but non-personalized models. Finally, we compare \textsc{ccm} to two baseline click models under different conditions.

\subsection{Fit to Real-World Data}
\label{sec:experiment1}

To study how well the \textsc{ccm} and \textsc{tcm} model the user behavior, we fit them to an existing dataset of real user interactions. The dataset was collected by \citet{jannach2021exploring} to assess the effect of different design choices on human decision-making. It contains $n = 776$ instances of clicks on recommendations presented in two settings, ranked list and carousels, with a comparable number of clicks in each setting. Despite the small size, we found this dataset to be the only publicly available data source of human interaction with a carousel-based interface, which can be used to validate the \textsc{ccm}.

The dataset has two parts. In both parts, the recommended items are presented in $m = 5$ rows and $K = 4$ items per row. In the first part (conditions $1$ to $4$), the items are presented as a single ranked list, row by row. We hypothesize that the user scans these items row by row, from left to right, and clicks on the first attractive item. We call these data \emph{ranked list}. The second part of the data (conditions $5$ to $8$) is similar to the first except that each row is labeled with the topic of items in that row, such as \say{Action Movies}. This can be viewed as a list of carousels. We hypothesize that the user scans the carousels from top to bottom and stops at the first attractive topic. After that, they examine the items within that carousel from left to right and click on the first attractive item. We call these data \emph{carousel}.

\begin{figure}[!h]
  \centering
  \includegraphics[width=0.5\textwidth]{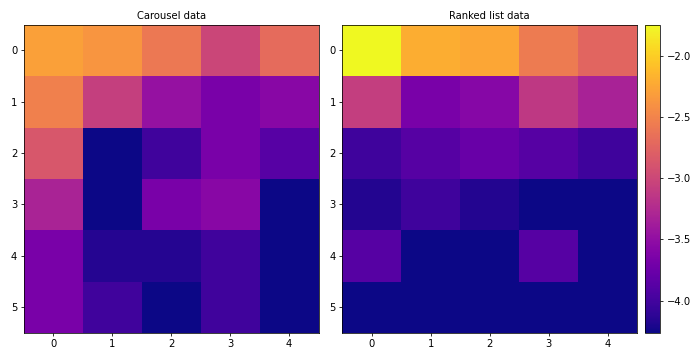}
  \caption{Empirical click probabilities for carousel (left) and ranked list (right) data.}
  \label{fig:exp1-real}
\end{figure}

In both the ranked list and carousel data, we compute empirical click probabilities for all positions and plot their logarithm in \cref{fig:exp1-real}. A closer look at these probabilities reveals different user interaction patterns in the two settings. In the carousel data, we observe more clicks in the first column, which represents the first items in all carousels. This is in contrast to ranked list data, where clicks are more concentrated in the first row, which represents the highest positions in the ranked list. The difference between the interactions is consistent with our proposed mathematical models, and we provide more quantitative evidence below.

One challenge in evaluating our mathematical models is that most items in our dataset appear only once. Therefore, it is impossible to accurately estimate their attraction probabilities. However, we know that the items are recommended in the order of decreasing relevance. Therefore, we parameterize the attraction probabilities of the items as follows. In the TCM, the click probability at position $(i, j)$ is computed as in \eqref{eq:tcm position click probability}, where $k = K (i - 1) + j$ and $p_{A_k} = p_0 \gamma^{k - 1}$. Here $p_0 \in [0, 1]$ is the highest attraction probability and $\gamma \in [0, 1]$ is its discount factor. Note that the attraction probability decreases with the rank of the item in the list, which is presented as a matrix. We denote the click probability at position $(i, j)$ by $\mu_\textsc{tcm}(i, j; p_0, \gamma, p_q)$, with parameters $p_0$, $\gamma$, and $p_q$. In the CCM, the click probability at position $(i, j)$ is computed as in \eqref{eq:ccm position click probability}, where $p_{A_{i, j}} = p_0 \gamma^{i + j - 2}$ and all parameters are defined as in the TCM. Again, the attraction probability decreases with the number of rows and columns, and we denote the click probability at position $(i, j)$ by $\mu_\textsc{ccm}(i, j; p_0, \gamma, p_q)$.

Let $\hat{\mu}_\textsc{list} \in [0, 1]^{m \times K}$ and $\hat{\mu}_\textsc{carousels} \in [0, 1]^{m \times K}$ denote the matrices of empirical click probabilities estimated from the ranked list and carousel data (\cref{fig:exp1-real}), respectively. For each model $M \in \set{\textsc{tcm}, \textsc{ccm}}$ and dataset $\cD \in \set{\textsc{list}, \textsc{carousels}}$, we compute
\begin{align*}
  \delta_{M, \cD}
  = \frac{1}{2} \min_{p_0, \gamma, p_q \in [0, 1]^3}
  \normw{\mu_M(\cdot, \cdot;  p_0, \gamma, p_q) - \hat{\mu}_\cD}\,,
\end{align*}
the minimum total variation distance between the click probabilities of the optimized model and their empirical estimates from the dataset. We approximate the exact minimization over $[0, 1]^3$ by grid search, where the grid resolution is $0.01$. We report all total variation distances in \cref{tab:exp1}.


\begin{table}[!h]
\begin{tabular}{l|ccccc}
\hline
\multicolumn{1}{c|}{Model} & {Data} & $p_0$ & $\gamma$ & $p_q$ & $\delta$      \\ \hline
\textsc{tcm} & Ranked list  & 0.17  & 0.92              & 0.02  & \textbf{0.086} \\
\textsc{ccm} & Ranked list  & 0.17  & 0.9               & 0.02  & 0.095          \\ \hline
\textsc{tcm} & Carousel & 0.099 & 0.96              & 0.01  & 0.133          \\
\textsc{ccm} & Carousel & 0.11  & 0.84              & 0.01  & \textbf{0.128} \\ \hline
\end{tabular}
\caption{Comparing the total variation distance between \textsc{tcm} and \textsc{ccm} models on ranked list and carousel data.}
\label{tab:exp1}
\end{table}

Our results in \cref{tab:exp1} show that the \textsc{tcm} fits the ranked list data better (smaller total variation distance $0.086$) than the \textsc{ccm} (larger total variation distance $0.095$). They also show that the \textsc{ccm} fits the carousel data better (smaller total variation distance $0.128$) than the TCM (larger total variation distance $0.133$). In summary, our mathematical models of click probabilities in the \textsc{tcm} and \textsc{ccm} match the observed clicks. We also use the results of this experiment to set the value of the termination probability $p_q$ in the remaining experiments. This value is $p_q = 0.01$.

\subsection{Generalization of \textsc{ccm}} 
\label{sec:generalization_of_cmm}

This experiment shows that the \textsc{ccm} generalizes to an unseen test set. Specifically, we show that the optimal recommendation under the \textsc{ccm} in the training set also has a high value on the test set.

This experiment is conducted on the MovieLens 100K dataset \cite{movielense_10.1145/2827872}, which consists of $100\,836$ ratings applied to $9\,724$ movies (items) in $19$ genres (topics) by $610$ users. For simplicity, we assume that each movie is associated only with one genre. For a movie with more than one genre, we assign the genre with the highest popularity among all users. The recommended movies in \textsc{ccm} are organized in $19$ carousels. Each carousel represents a genre and has a label that shows the topic of the carousel, such as \say{Action Movies}. We denote by $n_u$ the number of users and by $n_a$ the number of items. 

We randomly split the dataset into two equal sets: training set $\hat{\cD}$ and test set $\cD$. Then, for all users and items, we estimate the ratings using matrix factorization, which is a standard approach for rating estimation in recommender systems \cite{koren2009matrix}. We denote the estimated rating of item $a \in [n_a]$ by user $u \in [n_u]$ in the training (test) set by $\hat{r}_{u, a}$ ($r_{u, a}$). Next, we apply a softmax transformation to the estimated ratings (mean=$3.51$, standard deviation=$0.27$) and convert them into attraction probabilities (mean =$2\mathrm{e}{-4}$, standard deviation=$6.11\mathrm{e}{-5}$) in both the training and test sets,
\begin{align*}
  \hat{p}_{u, a}
  = \frac{\exp[\hat{r}_{u, a}]}{\sum_{a = 1}^{n_a} \exp[\hat{r}_{u, a}]}\,, \quad
  p_{u, a}
  = \frac{\exp[r_{u, a}]}{\sum_{a = 1}^{n_a} \exp[r_{u, a}]}\,.
\end{align*}
This is just a monotone transformation that transforms the estimated ratings of each user into a probability vector.

We evaluate the \textsc{ccm} as follows. Let $\hat{P}_\textsc{ccm}(A, u)$ ($P_\textsc{ccm}(A, u)$) be the click probability on recommendation $A$ by user $u$ on the training (test) set, computed using \eqref{eq:ccm click probability} and attraction probabilities $\hat{p}_{u, a}$ ($p_{u, a}$). First, we compute the best recommendation for user $u$ on the training set $\hat{A}_u = \argmax_A \hat{P}_\textsc{ccm}(A, u)$, where the maximization is done as described in \cref{sec:optimal solution}. Second, we evaluate $\hat{A}_u$ on the test set, by computing the test click probability $P_\textsc{ccm}(\hat{A}_u, u)$. Third, we compute the best recommendation for user $u$ on the test set $A_{u, *} = \argmax_A P_\textsc{ccm}(A, u)$, where the maximization is done as described in \cref{sec:optimal solution}. Finally, we compare the average test click probabilities, for all users $u$, of the best training and test recommendations, formally given by
\begin{align*}
  \frac{1}{n_u} \sum_{u = 1}^{n_u} P_\textsc{ccm}(\hat{A}_u, u)\,, \quad
  \frac{1}{n_u} \sum_{u = 1}^{n_u} P_\textsc{ccm}(A_{u, *}, u)\,.
\end{align*}

\begin{figure}[!h]
\centering
            \includegraphics[width=\textwidth/2]{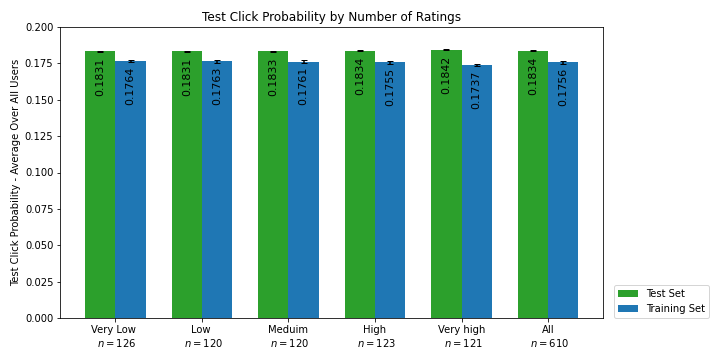}
            \caption[]%
            {{\small Average test click probabilities of the best recommendations on the training and test sets.}}    
            \label{fig:compare_optimal}
 \end{figure}

Our results are shown in \cref{fig:compare_optimal}. In addition to reporting the average click probability over all users, we break it into five groups based on the number of ratings per user: \say{very low} to \say{very high}. We choose the number of ratings because it represents the size of the user profile. The results of this experiment confirm that \textsc{ccm} can generate comparable recommendations to the best recommendations on the test set, both overall ($4.34\%$ difference) and in the five user groups ($3.0$-$5.8\%$ difference). This testifies to the generalization ability of our proposed model. Particularly, we show that our model does not overfit and performs well on the test set.

\subsection{Different Levels of Personalization in \textsc{ccm}}
\label{sec:personalization_in_cmm}

This experiment investigates the effect of personalization on the click probability of recommendations. We compare the click probability of recommendations generated by the \textsc{ccm} using personalized and two non-personalized attraction probabilities.

The setup of this experiment is the same as in \cref{sec:generalization_of_cmm}. The only difference is in the definitions of ratings in the training set. This definition affects how the optimal recommendation $\hat{A}_u$ is chosen, but not how it is evaluated. That is, $P_\textsc{ccm}(\hat{A}_u, u)$ is the value of $\hat{A}_u$ under the personalized ratings on the test, as defined in \cref{sec:generalization_of_cmm}. The first approach, \textsc{personalized}, uses the definition of training set ratings in \cref{sec:generalization_of_cmm}. The second approach, \textsc{popular}, calculates the rating of item $a$ for user $u$ as
\begin{align*}
  \hat{r}_a
  = \frac{\sum_{u = 1}^{n_u} \hat{r}_{u, a}}{n_u}\,.
\end{align*}
This is not personalized because the rating of item $a$ is the same for all users. The last approach, \textsc{random}, assigns random ratings $\hat{r}_a \in [1, 5]$ to all items.

\begin{figure}[!h]
\centering
            \includegraphics[width=\textwidth/2]{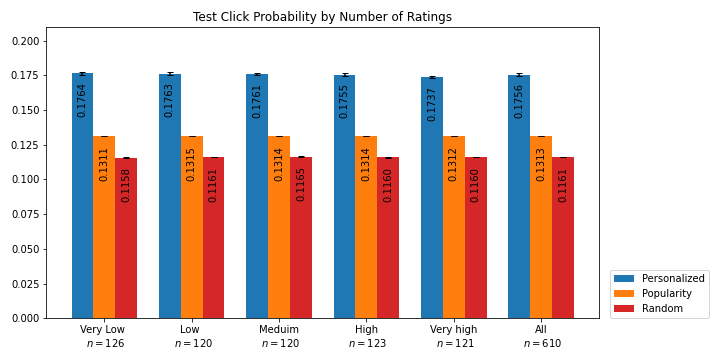}
            \caption[]%
            {{\small Comparing the click probability of \textsc{CCM} in different settings.}}    
            \label{fig:compare_settings}
 \end{figure}

Our results are reported in \cref{fig:compare_settings}. In addition to reporting the average click probability over all users, we break it over the same five user groups as in \cref{fig:compare_optimal}. As explained before, this segmentation helps to better visualize the effect of user profile size on the click probability. We observe that the personalized model outperforms the non-personalized baselines by a large margin. Specifically, the average click probability over all users in \textsc{popular} decreases by $25\%$ from that in \textsc{personalized}; and the average click probability over all users in \textsc{random} decreases by $34\%$ from that in \textsc{personalized}. These improvements are consistent across all five user groups, which indicates stability with respect to the profile size. We conduct this experiment because the non-personalized baselines depend less on the training data of a given user, and thus may generalize better to the test set. This experiment shows that this is not the case.

\subsection{Comparison of \textsc{ccm} to Other Click Models}
\label{sec:comparison_with_others}

\begin{figure}[h!]
\centering
            \includegraphics[width=\textwidth/2]{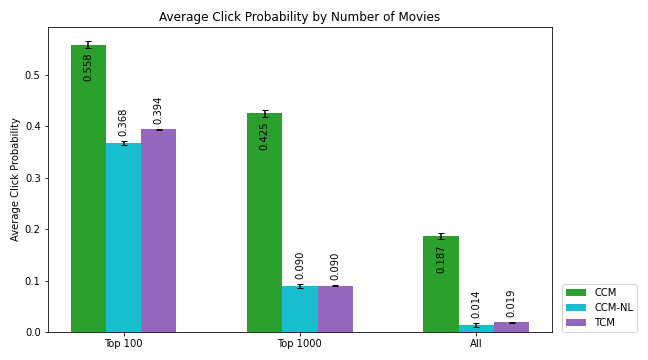}
            \caption[]%
            {{\small Comparing the average click probability in CCM, CCM-NL and TCM.}}    
            \label{fig:compare_models}
 \end{figure}

\begin{figure*}
\centering
            \includegraphics[width=\textwidth]{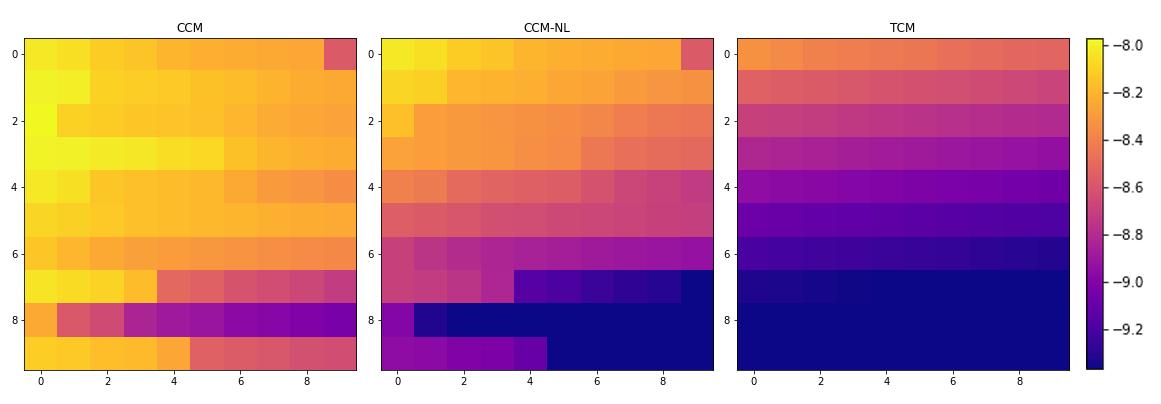}
            \caption[]%
            {{\small Sample distribution of log click probability in CMM, CMM-NL and TCM for a random user. Darker colors mean less click probability.}}    
            \label{fig:exp2-heatmap}
 \end{figure*}

This experiment compares click probabilities under the \textsc{ccm} with two other click models: \textsc{tcm} and a variant of \textsc{ccm} called \textsc{ccm-nl}. The goal of this experiment is to show the gain in click probabilities when using the \textsc{ccm}.

\textsc{tcm} (\cref{sec:terminating cascade model}) is a cascade click model \cite{craswell08experimental} that models a user that may terminate unsatisfied. \textsc{ccm-nl}, which stands for a \emph{carousel click model with no labels}, is a variant of the \textsc{ccm} that models the scenario when topic labels are removed. Specifically, we compute the optimal list using the \textsc{ccm} and then remove the labels of the carousels. This means that the user browses the recommendations as if they were a single ranked list, and we adopt the \textsc{tcm} for this simulation. We do this to study the importance of labels for carousels and to see to what extent they affect the average click probability in \textsc{ccm}. We use the same evaluation protocol as in \cref{sec:generalization_of_cmm}.

Our results are reported in \cref{fig:compare_models}. The click probabilities are reported for three different sets of items: top $100$, top $1\,000$, and all; where the items are ranked by the sum of their ratings. This experiment shows that the click probability in the \textsc{ccm} is significantly higher than in the \textsc{tcm} and \textsc{ccm-nl}. Specifically, when recommending top $100$ items, the click probability in \textsc{tcm} decreases by $29.4\%$ from that in \textsc{ccm}; and the click probability in \textsc{ccm-nl} decreases by $34.0\%$ from that in \textsc{ccm}. The improvement increases as the number of items increases. Specifically, when recommending all items, the click probability in \textsc{tcm} decreases by $89.8\%$ from that in \textsc{ccm}; and the click probability in \textsc{ccm-nl} decreases by $92.5\%$ from that in \textsc{ccm}. In summary, the \textsc{ccm} attracts about $10$ times as many clicks as both the \textsc{tcm} and \textsc{ccm-nl} when recommending all items. This trend in improvement indicates that \textsc{ccm} is a good candidate for practice, where a large number of recommended items is common.

\subsection{More Realistic \textsc{ccm}}

The \textsc{ccm} assigns the same termination probability to all items in the carousel. However, in practice, the items that are initially not displayed are less likely to be examined, because the user needs to scroll to see them. To study this behavior, we assign different termination probabilities to different parts of the carousel: $p_q = 0.01$ to the first $10$ columns and $p_q = 0.1$ to the rest. Selecting the first $10$ items as the visible part of the carousel is an intuitive choice based on real-life recommender systems. Other than this, the setting is the same as in \cref{sec:comparison_with_others}.

\begin{figure}[H]
\centering
            \includegraphics[width=\textwidth/2]{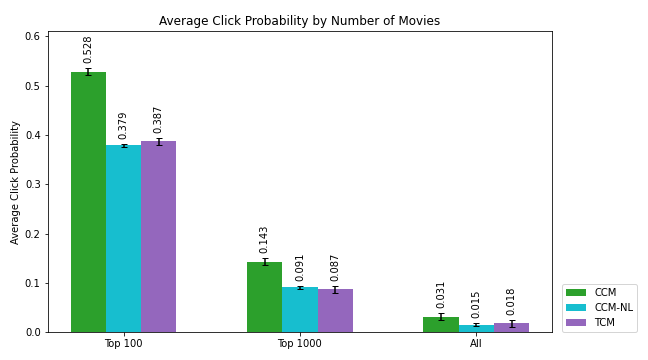}
            \caption[]%
            {{\small Comparing the average click probability in a more realistic CCM, CCM-NL and TCM.}}
            \label{fig:compare_models_hidden}
\end{figure}

\cref{fig:compare_models_hidden} compares the \textsc{ccm}, \textsc{ccm-nl}, and \textsc{tcm} in the new setting. We observe that the performance of \textsc{ccm} worsens. In absolute terms, the click probability for top $100$ items is comparable to that in \cref{fig:compare_models}. However, for all items, it is about $5$ times lower. Relatively to the baselines, when recommending top $100$ items, the click probability in \textsc{tcm} decreases by $26.7\%$ from that in \textsc{ccm}; and the click probability in \textsc{ccm-nl} decreases by $28.23\%$ from that in \textsc{ccm}. When recommending all items, the click probability in \textsc{tcm} decreases by $41.9\%$ from that in \textsc{ccm}; and the click probability in \textsc{ccm-nl} decreases by $51.6\%$ from that in \textsc{ccm}. Although the improvement for all items is not as impressive as in \cref{fig:compare_models}, \textsc{ccm} still outperforms both baselines by a healthy margin.

\subsection{Visualizing the Log Click Probability}

To visualize the reason behind the considerably better performance in the \textsc{ccm}, we plot the average log click probability for all users using \eqref{eq:ccm position click probability} in the first ten rows and columns of the optimal recommendation in \textsc{ccm} and the two baselines in \cref{fig:exp2-heatmap}. The light color (yellow) in the plots corresponds to high average click probabilities while the dark color (blue) corresponds to low average click probabilities. In \textsc{ccm}, we observe that the items at the beginning of each row (the first few items in each carousel) are more likely to be clicked by the user. We can also observe the effect of removing labels from the carousels in \textsc{ccm-nl}. This is manifested by overall darker colors because the user struggles to find the carousel with attractive items and leaves unsatisfied. The last plot shows that the average click probability in \textsc{tcm} decreases uniformly. This is expected when the recommended items are ranked in the order of decreasing relevance.

\section{Conclusions}
\label{sec:conclusions}

Carousel-based recommendation interfaces have become a de-facto standard approach to recommending items to end users in many real-life recommenders. From an analytical point of view, their benefits are not well understood. In this work, we propose a carousel click model where the user examines carousels proportionally to the attraction of items in them, and then clicks on the items within the examined carousel proportionally to their individual attraction. This model is motivated by the cascade model in a single ranked list. We compare these models analytically and show that the user is more likely to click in the carousel click model because a structured examination of a large item space is more efficient than scanning a single ranked list. These observations are supported by experiments on real data, where we also show how to integrate the carousel click model with a recommender based on matrix factorization.

There are many directions for future work. First, while working on this topic, we noted that there is no large-scale public dataset of how users interact with carousels. To the best of our knowledge, the small dataset in \cref{sec:experiment1} is the largest such dataset, and even this one could not be fully used to validate our model. Therefore, most of our experiments are synthetic. Based on our experience, we believe that it is critical to publicly release a large-scale carousel dataset to encourage more research in this area, similarly to what the Yandex dataset \citep{yandex} did for the regular click models \citep{chuklin15click}. Second, our carousel click model is only one potential model for carousels, motivated by the cascade model in ranked lists. An alternative model could be based on the position-based model, where the user would examine the item at position $(i, j)$ proportionally to the probability of examining row $i$ and column $j$. We believe that many such models could be developed in the future, similarly to the countless click models for ranked lists \citep{chapelle09dynamic}. Finally, we take a model-based approach in this work, where we first learn a model of clicks and then make the best recommendation under it. An alternative in a single ranked list is learning to rank directly from clicks \cite{joachims2002optimizing}.  How to do that with carousels is unclear and we hope that our paper will motivate some work in this area.

\bibliographystyle{ACM-Reference-Format}
\bibliography{refs}  

\end{document}